\documentclass[sigconf,nonacm]{acmart}

\copyrightyear{2024} 

\acmYear{2024} 

\setcopyright{rightsretained} 

\acmConference[SA Technical Communications '24]{SIGGRAPH Asia 2024 Technical Communications}{December 3--6, 2024}{Tokyo, Japan}

\acmBooktitle{SIGGRAPH Asia 2024 Technical Communications (SA Technical Communications '24), December 3--6, 2024, Tokyo, Japan}\acmDOI{10.1145/3681758.3697987}

\acmISBN{979-8-4007-1140-4/24/12}

\usepackage{color}
\usepackage{tabularx}
\usepackage{multirow}
\usepackage{subcaption}

\newcolumntype{Y}{>{\centering\arraybackslash}X}

\newcommand{\copylabel}[2]{\textit{\footnotesize #1 $\copyright$ #2}}

\acmSubmissionID{tcom\_118}

\begin{document}

\title{Eyelid Fold Consistency in Facial Modeling}

\author{Lohit Petikam}
\email{lohitpetikam@microsoft.com}
\orcid{0000-0001-6629-7490}
\affiliation{%
  \institution{Microsoft}
  \city{Cambridge}
  \country{United Kingdom}
}
\author{Charlie Hewitt}
\email{chewitt@microsoft.com}
\orcid{0000-0003-3943-6015}
\affiliation{%
  \institution{Microsoft}
  \city{Cambridge}
  \country{United Kingdom}
}
\author{Fatemeh Saleh}
\email{fatemehsaleh@microsoft.com}
\orcid{0000-0002-3695-9876}
\affiliation{%
  \institution{Microsoft}
  \city{Cambridge}
  \country{United Kingdom}
}
\author{Tadas Baltru\v{s}aitis}
\email{tabaltru@microsoft.com}
\orcid{0000-0001-7923-8780}
\affiliation{%
  \institution{Microsoft}
  \city{Cambridge}
  \country{United Kingdom}
}

\renewcommand{\shortauthors}{Petikam et al.}

\begin{abstract}
Eyelid shape is integral to identity and likeness in human facial modeling.
Human eyelids are diverse in appearance with varied skin fold and epicanthal fold morphology between individuals.
Existing parametric face models express eyelid shape variation to an extent, but do not preserve sufficient likeness across a diverse range of individuals.
We propose a new definition of eyelid fold consistency and implement geometric processing techniques to model diverse eyelid shapes in a unified topology.
Using this method we reprocess data used to train a parametric face model and demonstrate significant improvements in face-related machine learning tasks.
\end{abstract} 




\maketitle

\section{Introduction}

\begin{figure}
    \centering
    \includegraphics[width=\linewidth]{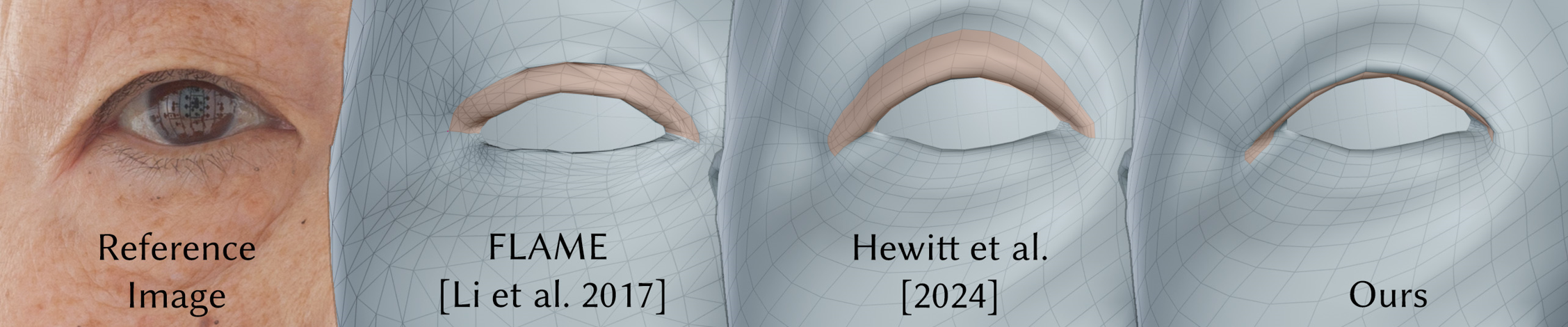}\\
    \caption{For an individual with a hooded eyelid and epicanthal fold; FLAME~\cite{FLAME:SiggraphAsia2017} cannot model either in geometry. \citet{hewitt2024look} do not model hoodedness causing an over-smoothed eyelid crease. Our definition ensures eyelid creases are sharp and that hoodedness and epicanthal folds are explicitly modeled in geometry. Orange depicts the surface beneath the eyelid crease. \copylabel{Photo}{Triplegangers}}
    \label{fig:eyelid-inconsistency}
\end{figure}
\begin{figure}
    \centering
    \includegraphics[width=\linewidth]{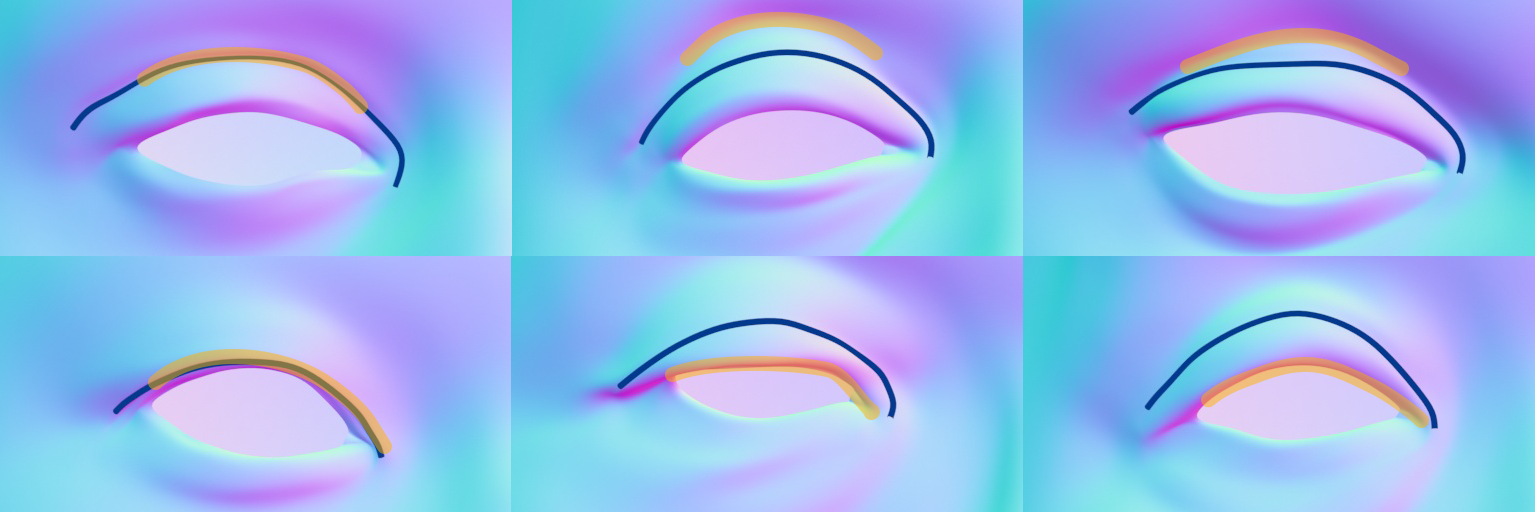}\\ 
    \caption{Topological inconsistencies in eyelids of training data of \citet{hewitt2024look}. The same geometric feature (yellow) is not always represented by the same edge-loop (blue).}
    \label{fig:old-inconsistency}
\end{figure}

Eyes are a key part of human identity and likeness that are critical for communication.
Specifically, the shape 
created by the folding of the eyelid is an essential feature for distinguishing between individuals.
Eyelid folds are highly correlated with protected attributes like ethnicity and age and, given their importance for likeness, are critical to consider in computer graphics and vision applications.
The risk of misrepresenting a person's eye shape is significant for the fairness and inclusivity of any software.
The issue of accurate eye representation is particularly relevant to human facial modeling, in which face shape can be described using a parametric or statistical model.
Such models are useful for common computer vision tasks such as face tracking and reconstruction.
Existing parametric face models are typically based on large datasets of 3D human face scans \cite{FLAME:SiggraphAsia2017,paysan20093d,wood2021fake}.
While their training data may be reasonably well balanced across protected attributes, these models do not represent a diverse range of eyelid shapes at high quality, as shown in Figure~\ref{fig:eyelid-inconsistency}.
Given the seemingly adequate \emph{quantity} of training data for these models, this raises questions about the \emph{quality} of the data being used.
We demonstrate that this failure of inclusive representation of eye shape in existing models is primarily due to a lack of consistency in their training data.
Figure~\ref{fig:old-inconsistency} shows this inconsistency for some training samples of \citet{hewitt2024look}.

Defining what consistency means for eyelid folds across all of humanity is clearly a difficult task.
A good definition suited to graphics applications should 
(1) allow for semantically meaningful interpolation between different shapes 
(2) support a common texture space 
(3) allow for application of common animations 
(4) have a fold corresponding to a fixed set of vertices / edge loop.

We introduce an objective and consistent definition of eyelid shape across humanity which satisfies the above properties.
We apply this to reprocess existing head scan data and retrain the parametric face model of \citet{hewitt2024look}, showing significantly improved representation of diverse eyelid folding.
We evaluate the impact of the retrained model on downstream computer vision tasks to validate its potential for real-world applications. 

\section{Describing Eyelid Shape}

\begin{figure}
    \centering
    \footnotesize
    \begin{tabularx}{0.9\linewidth}{YYY}
         Open & Closing & Proposed Annotation
    \end{tabularx}
    \includegraphics[width=0.9\linewidth]{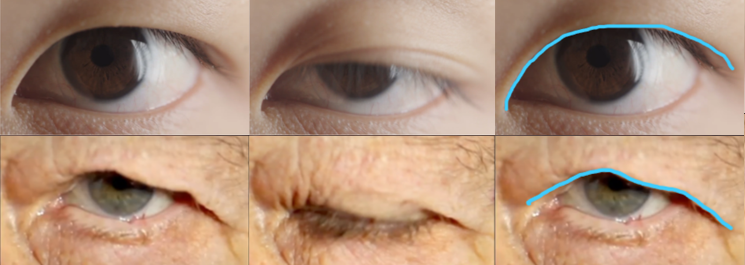}\\
    \caption{Left: Hooded and monolid eyelids without a prominent eyelid fold. Middle: Partially closed revealing the fold. Right: Our annotation acknowledging that the fold is just behind the eyelid hood. \copylabel{Videos}{AILA\_IMAGES, Damato / Adobe Stock}}
    \label{fig:real-eyelids}
\end{figure}

Human eyelid morphology is complex, as shown by \citet{bermano2015detailed} who precisely track the occluded surface of skin due to the upper half of the upper eyelid overlapping the lower half.
3D morphable models, however, cannot track eyelids with such precision due to their fixed, lower-fidelity mesh topology. 
Existing face models \cite{paysan20093d,FLAME:SiggraphAsia2017,wood2021fake,hewitt2024look} do not enforce consistency around eyelids and rely on texture to recover details. 
This is insufficient for meaningful interpolation, and makes textures and animation blendshapes for different face shapes incompatible.
\citet{wen2017real} use a face model with eyelid consistency, but their consistency definition excludes hooded eyelids with epicanthal folds. 
In reality, the eyelid fold of hooded and monolid eyelids is occluded behind the skin falling in front of the eye and only becomes visible when these individuals blink, raise their eyebrows, or look around. 
In Figure~\ref{fig:real-eyelids} we show hooded and monolid eyelids without prominent folds when open, and folds becoming visible when while closing.


\begin{figure}
    \centering
    \footnotesize
    \begin{tabularx}{\linewidth}{YYYY}
         Hooded (outer side) & Partially hooded & Non-hooded & Hooded with epicanthal fold
    \end{tabularx}
    \includegraphics[width=\linewidth]{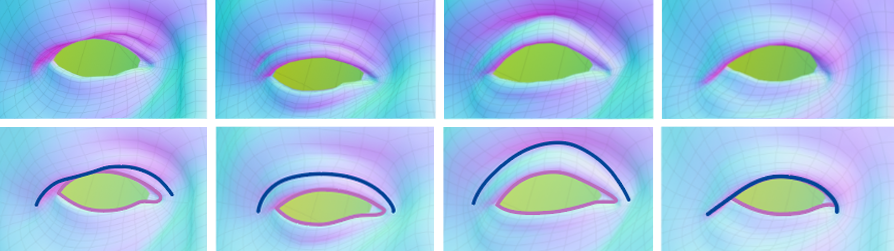}\\
    \caption{Eyelid shapes (top) with annotations (bottom) based on our definition of eyelid consistency.}
    \label{fig:consistency-def}
\end{figure}

We propose a new eyelid consistency definition that is inclusive of hooded eyelids and epicanthal folds. 
In our consistency definition we partition the eyelid into two parts, the upper and lower lid, separated by a crease that is represented with the same edge-loop across all meshes. 
For non-hooded eyes, both the upper and lower lids are visible with minimal overlap over each other. 
For hooded eyes, only the upper lid is visible with the crease forming the bottom of the visible eyelid, and the lower lid geometry tucked behind the upper lid, as is physically the case. 
See Figure~\ref{fig:consistency-def} for annotations of this crease on varying eye shapes.

\section{Data Processing}

To impose eyelid consistency on the training data of \citet{hewitt2024look} we retopologise all scans to represent diverse eyelid shapes consistently and with explicitly modeled folds.
We retrain the face model of \citet{hewitt2024look} using these improved retoplogisations.

\paragraph{Eyelid topology templates}
The face model training data of \citet{hewitt2024look} are the result of registering a base-mesh with desired topology to a diverse set of human head scans using commercial software \cite{Wrap3}. 
The process to retopoligise a scan into a common topology requires correspondence points to be manually annotated on the scan and the base-mesh, for example around the eyes, mouth and ears.
However, even when using our consistency definition, annotating eyelid correspondence points before retopologising yields artifacts around the eyelids unless the initial base-mesh is 
similar in eyelid shape to the scan.

\begin{figure}
    \centering
    \footnotesize
    \begin{tabularx}{\linewidth}{YYY}
         Non-hooded & Partially hooded & Hooded with epicanthal fold
    \end{tabularx}
    \includegraphics[width=\linewidth]{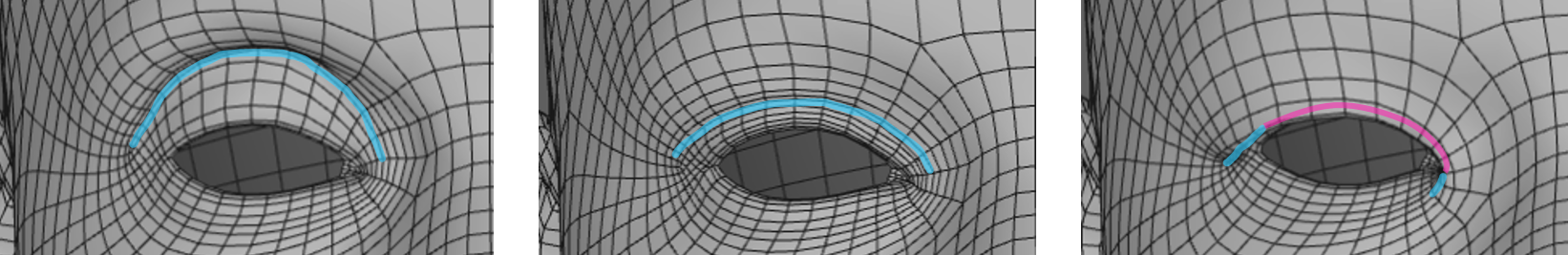}\\
    \caption{Template meshes in the topology of \citet{hewitt2024look} with explicitly modeled eyelid creases on shared edge-loop (cyan). The far-right template has an explicitly modeled crease just behind the eyelid hood and epicanthal fold (pink).}
    \label{fig:eyelid-templates}
\end{figure}

To achieve clean, representative retopologisations, we manually sculpt a set of diverse template eyelid shapes in the 3D topology of \citet{hewitt2024look} that explicitly model the eyelid fold with a common crease edge-loop for all eyelid types in our consistency definition, shown in Figure~\ref{fig:eyelid-templates}.
These template shapes help 
retain the desired explicitly-modeled eyelid crease and animation properties when 
retopologising the training scans. 
The template shapes also help 
faithfully reconstruct hooded eyelids, monolids, and epicanthal folds that are not well represented in the scan surface due to noise and limited resolution. 
By interpolating between the non-hooded, partial-hood, and hooded-epicanthal shapes, at different parts of the eyelid, we generate templates covering the eyelid diversity in our scan dataset.
Since the scan surface and texture are of insufficient quality to automatically determine a suitable interpolated template for each scan being retopologised, we create a custom tool to manually 
retopologise eyelids based on human judgment.

\begin{figure}
    \centering
    \includegraphics[width=\linewidth]{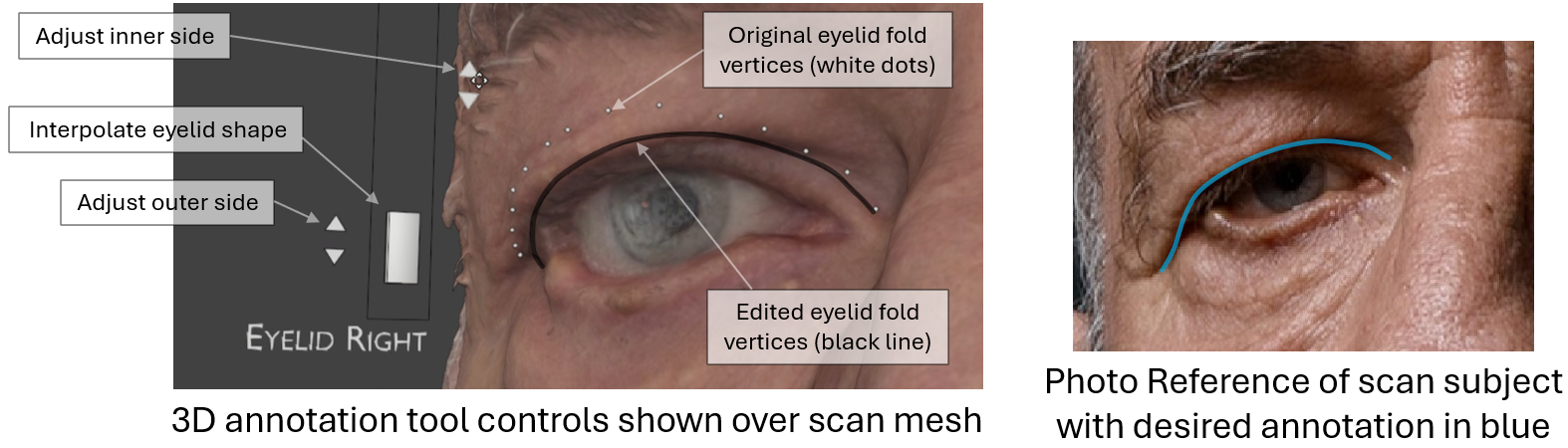}\\
    \caption{Custom 3D tool to annotate and retopologise eyelids of a face scan. \copylabel{Scan and photo}{Triplegangers}}
    \label{fig:eyelid-tool}
\end{figure}

\paragraph{Eyelid retopology tool}
To enable manual annotation we compute 20 uniformly-spaced interpolations between the three templates and retopologise the interpolations to all face scans in the dataset. 
This gives 20 potential retopologisations (retopos) for each face scan, each with the eyelid fold vertices at different locations on the eyelid. 
In the tool the user can smoothly interpolate between these with a slider to select the one that best conforms to our consistency definition. 
Users can adjust the inner and outer shape independently (within regions defined by vertex masks) using two additional sliders, see Figure~\ref{fig:eyelid-tool}. 
For eyelid shapes that cannot be reached through the tool we manually sculpt the retopos to represent the scan using the tools provided by Wrap \cite{Wrap3}.

\paragraph{Eyelid fold clarity and augmentation}
The training retopos of \citet{hewitt2024look} also have inconsistent eyelid fold clarity.
Many of them have blurry or overly smooth eyelid folds misrepresenting the physical fold of skin that is present. 
We ensure that eyelid folds are consistently represented by sharp creases in all retopos.
To achieve this we sculpt a custom blendshape that pinches the edge-loops surrounding the eyelid fold together, sharpening the eyelid crease.
In another custom tool, the blendshape strength and orientation are manually tweaked for each retopo to best match the scan without producing artifacts.
Having ensured consistent modeling of eyelid folds in the dataset, we augment the dataset with horizontally mirrored retopos to encourage the model to represent asymmetry in eyelid shape.
After this augmentation, we retrain the face model using the same method as \citet{hewitt2024look}.

\section{Results}
In this section we show how 3D model fitting and face shape reconstruction benefit from our retrained face model with enhanced eyelid consistency described above.

\paragraph{Eyelid shape metric}
We metricate eyelid shape directly in order to measure and evaluate diversity and accuracy of the shape model and downstream tasks.
In the previous sections, we described different eyelid shapes such as non-hooded, partially-hooded, or hooded, but the level of ``hoodedness'' can also vary at different parts of the eye. 
For example, older individuals can have eyelids that hood only on the outer side, while East-Asian individuals can have epicanthal folds covering just the inner eye corner. 
We define a metric that measures the degree of hoodedness at any point between the outer and inner sides of the eyelid. 
The hoodedness value at a point on the eyelid fold is a proportion of how far the eyelid crease is between the eyebrow and eyelid margin (where the eyelid meets the eyeball). 

\begin{figure}
    \centering
    \includegraphics[width=0.9\linewidth]{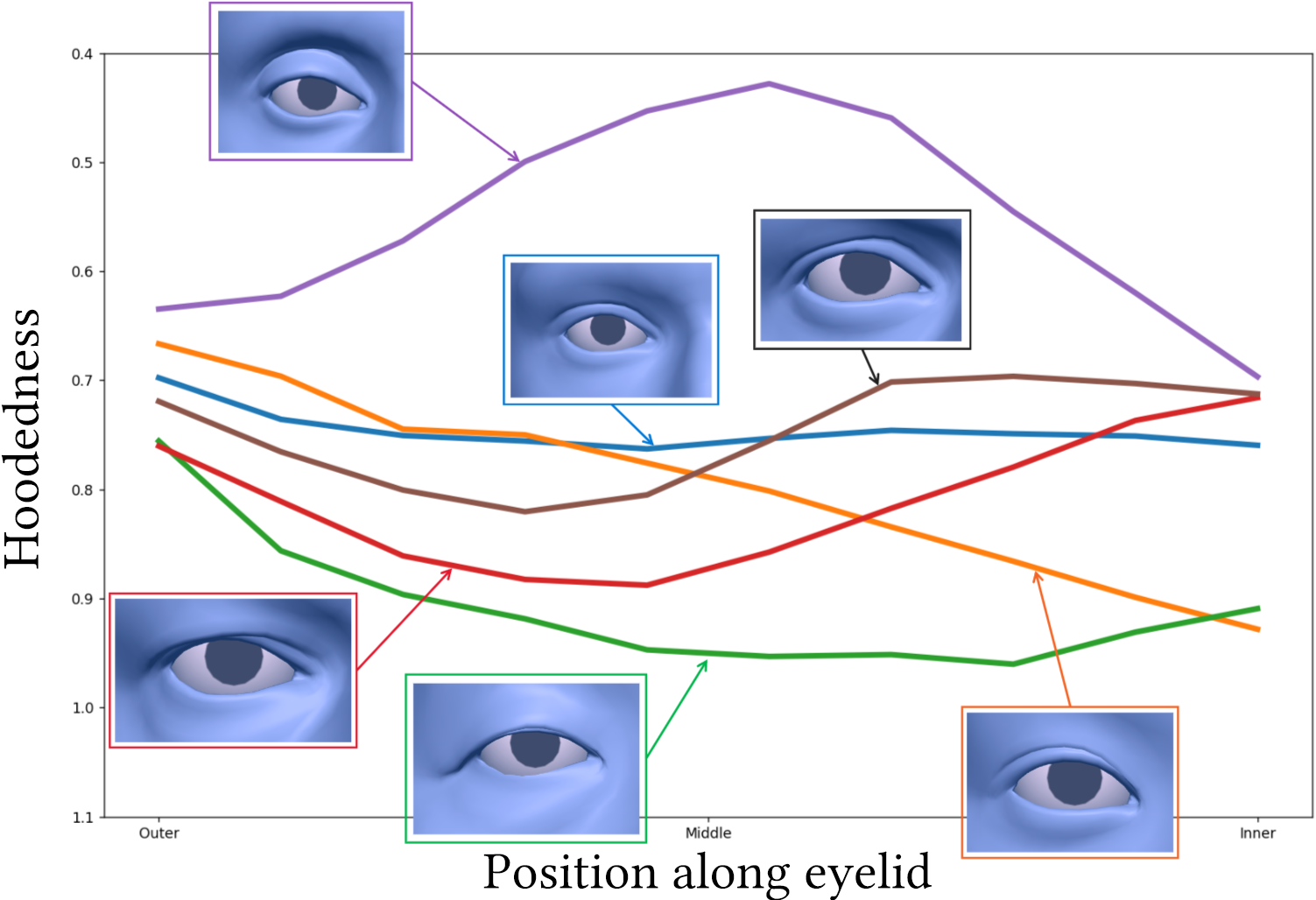}
    \caption{Eyelid shape profiles measured using our metric based on hoodedness, for a sample of our eyelid-consistent retopologizations of real scans.}
    \label{fig:eyelid-profiles}
\end{figure}

We measure this proportion by first determining curves $c_{\text{margin}}$,  $c_{\text{crease}}$,  $c_{\text{brow}}$, by sampling the eyelid margin, eyelid crease, and eyebrow vertices respectively. 
These curves are resampled by distance and projected to 2D as we only consider the frontal view. 
We define parameter $t\in[0,1]$ which denotes the temporal and nasal sides of the eyelid respectively.
For a given value of $t$, we define $v(t)= c_{\text{margin}}(t) - c_{\text{brow}}(t)$ and the hoodedness $h(t) = |proj_{\vec{v}}(c_{\text{crease}}(t) - c_{\text{brow}}(t))| / |v(t)|$ .
So $h=0.0$ denotes the brow line, and $h=1.0$  denotes the eyelid margin.
$h$ can exceed 1.0 where the eyelid fold passes in front the eyelid margin.
Example eyelid shape profiles measured using this metric are shown in Figure \ref{fig:eyelid-profiles}.

\begin{figure}
    \centering
    \footnotesize
    \begin{tabularx}{\linewidth}{YYY}
         Ground-Truth & \citet{hewitt2024look} & Ours
    \end{tabularx}
    \includegraphics[width=\linewidth]{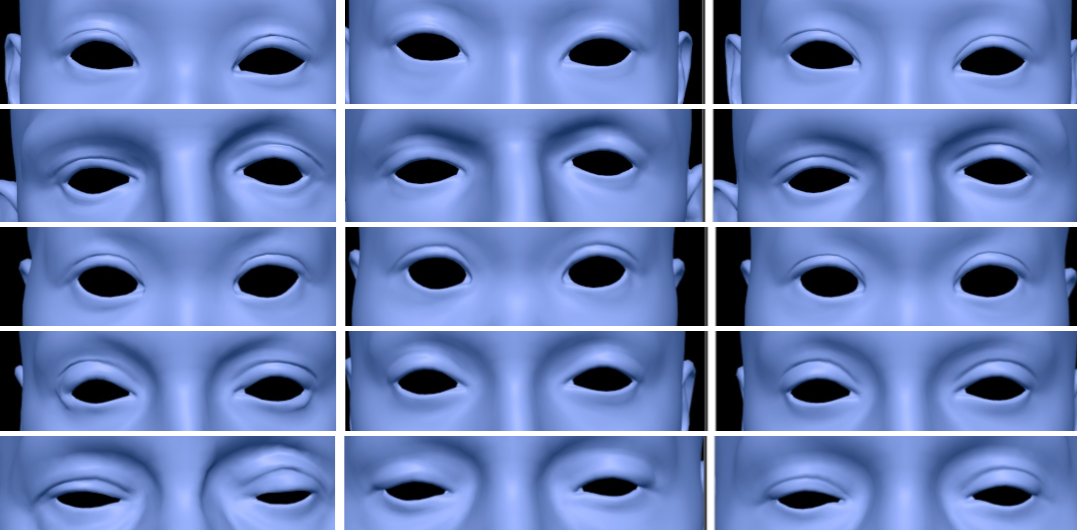}
    \caption{Comparison of fitting the face model of \citet{hewitt2024look} (middle) and our retrained model (right) to our eyelid-consistent ground-truth test retopos (left).}
    \label{fig:3d-fitting}
\end{figure}

\begin{figure}
    \centering
    \includegraphics[width=0.9\linewidth]{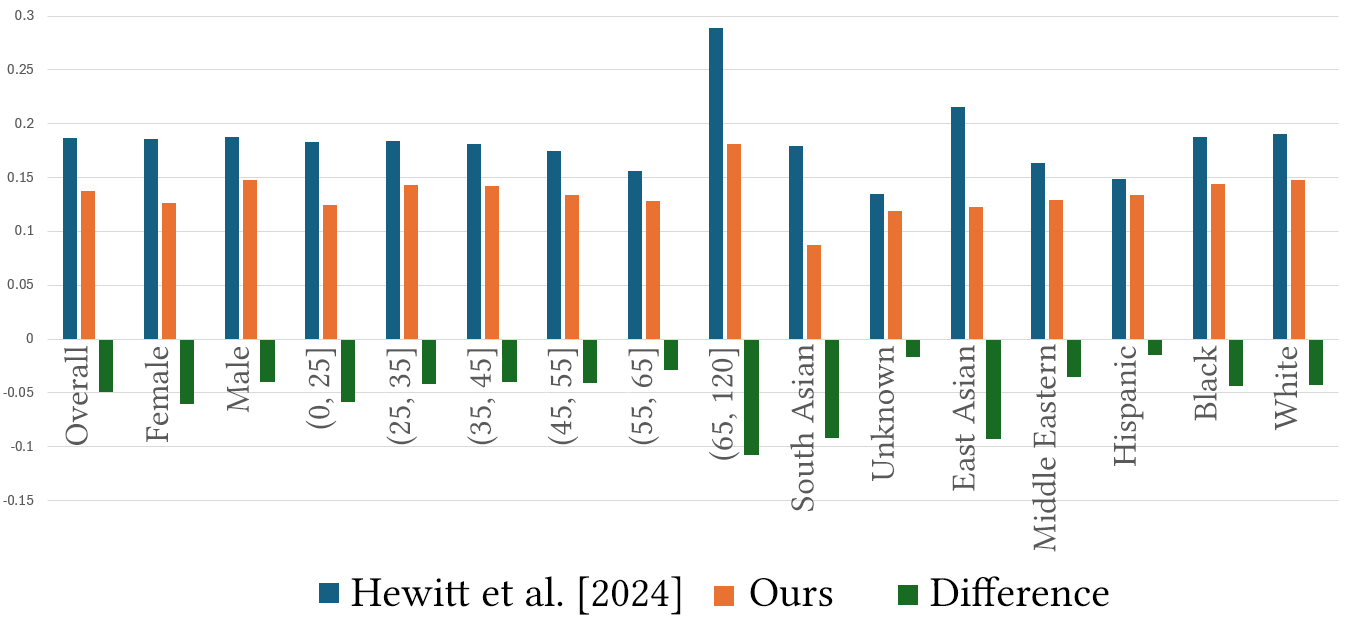}
    \caption{Eyelid shape error for fits to test set of 136 ground-truth retopoligised scans split by demographics. Our retrained model reconstructs more accurate eyelid shape than the model of \citet{hewitt2024look}.}
    \label{fig:3d-fit-eyelid-error}
\end{figure}

\paragraph{3D model fitting}
Our retrained face model provides greater visual likeness and quality in 3D test fits compared to the model of \citet{hewitt2024look}, using their 3D fitting procedure. 
Where the previous model produced faint and blurry eyelid creases, ours produces clearly defined creases that match in shape with the ground-truth (GT) retopo. 
Ours also more clearly models epicanthal folds that connect with the eyelid crease when present in the GT (Figure~\ref{fig:3d-fitting}).
Using our eyelid shape metric we quantify the accuracy of the model on the test set of \citet{hewitt2024look}.
Figure~\ref{fig:3d-fit-eyelid-error} shows lower eyelid shape error across all demographics.

\begin{figure}
    \centering
    \footnotesize
    \begin{tabularx}{\linewidth}{YYY}
         Input Photo & \citet{hewitt2024look} & Ours
    \end{tabularx}
    \includegraphics[width=\linewidth]{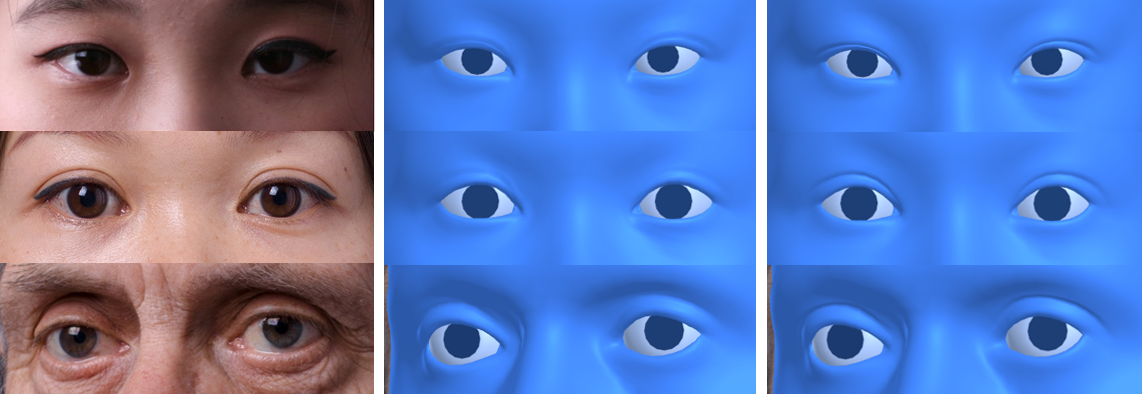}
    \caption{Comparison of fits to 2D landmarks using the model of \citet{hewitt2024look} and our retrained model. \copylabel{Photos}{\citet{stirling} and Triplegangers}}
    \label{fig:registration-compare}
\end{figure}

\begin{figure}
    \centering
    \begin{subfigure}{0.49\linewidth}
        \includegraphics[width=\linewidth]{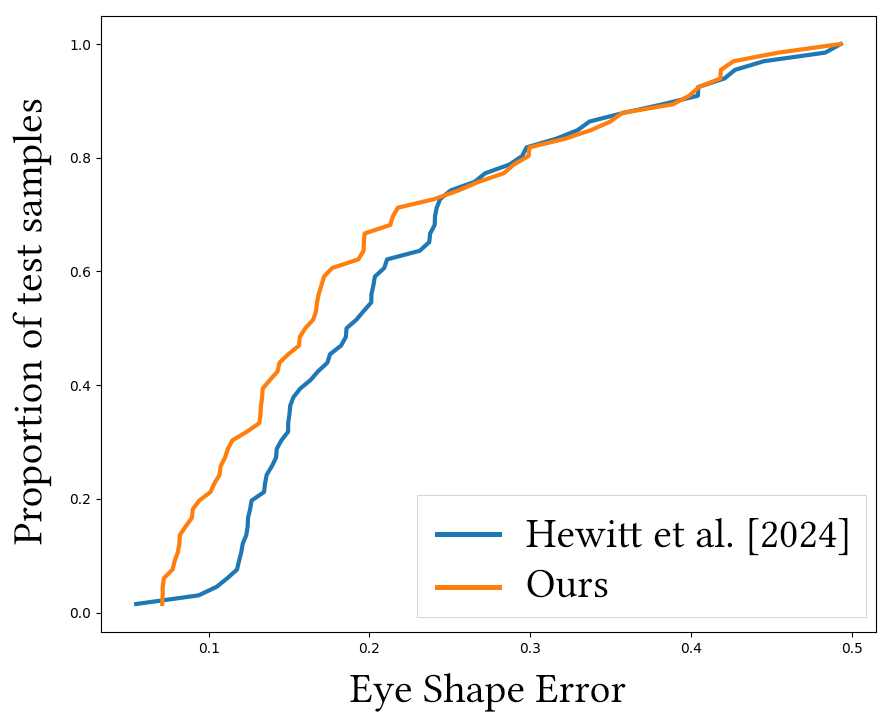}
        \caption{Landmark-based}
        \label{fig:reg-cdf}
    \end{subfigure}
    \hfill
    \begin{subfigure}{0.49\linewidth}
        \includegraphics[width=\linewidth]{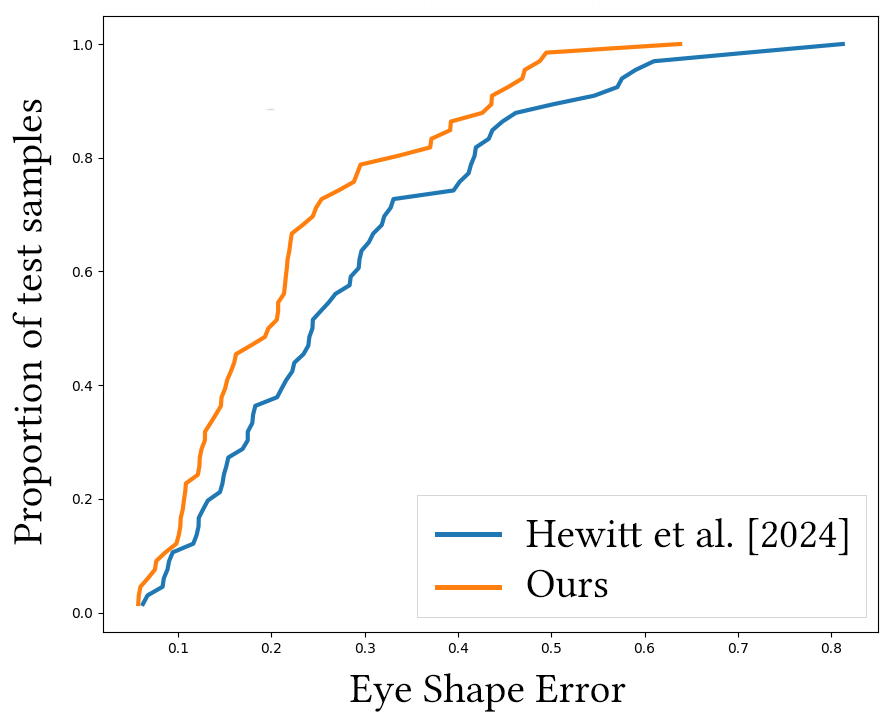}
        \caption{Neural}
        \label{fig:mifi-cdf}
    \end{subfigure}
    \caption{Improved cumulative error in eyelid shape when using our retrained model on test set of 66 individuals (toward top-left is better) for landmark-based (a) and neural (b) face shape reconstruction.}
\end{figure}

\paragraph{Landmark-based face shape reconstruction}
We evaluate the effect of our eyelid-consistent face model on landmark-based face shape reconstruction with model-fitting using the method of \citet{hewitt2024look}.
First, we replace the face model in their synthetic training data generation pipeline with ours in order to generate synthetic images of faces with diverse eyelid shapes. 
We then train a dense landmark regression DNN on this new synthetic dataset and use the same model fitting process with an updated identity prior based on our data. 
Figure~\ref{fig:registration-compare} shows that results using our retrained model reconstruct eyelid shape with more likeness to the input photo.
This is supported by the quantitative results shown in Figure~\ref{fig:reg-cdf}.

\begin{figure}
    \centering
    \footnotesize
    \begin{tabularx}{\linewidth}{YYY}
         Input Photo & \citet{hewitt2024look} & Ours
    \end{tabularx}
    \includegraphics[width=\linewidth]{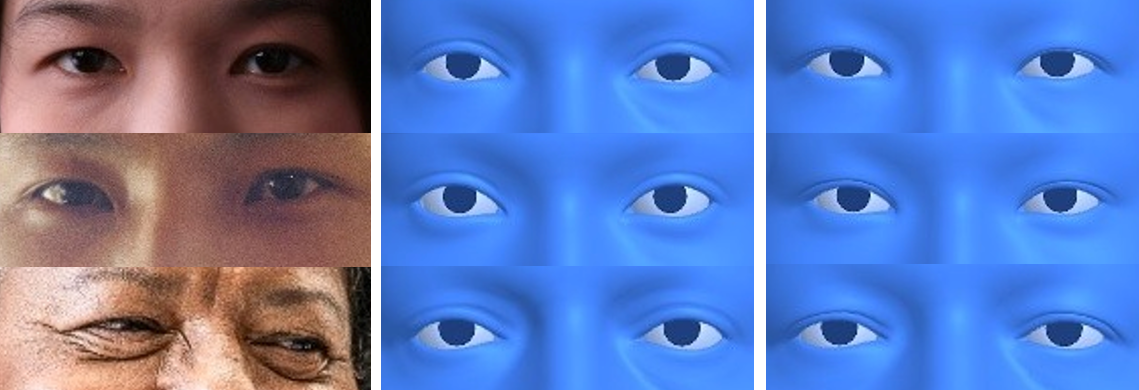}\\
    \caption{Comparison of neural face shape prediction using the model of \citet{hewitt2024look} and our retrained model. \copylabel{Photos}{\citet{stirling} and Microsoft}}
    \label{fig:mifi-compare}
\end{figure}

\paragraph{Neural face shape reconstruction}
We also evaluate the task of direct 3D face shape reconstruction with a neural network. 
We use the synthetic data described above to train a convolutional neural network to predict neutral face shape parameters of our face model directly from an input portrait photo \cite{zielonka2022towards, zhang2023accurate}. 
Figure~\ref{fig:mifi-compare} shows significantly more likeness is captured using our face model as a basis for neural face shape prediction. 
This is also supported by the quantitative results shown in Figure~\ref{fig:mifi-cdf}.

\section{Conclusion and Future Work}
We propose a new definition of consistency for modeling human eyelid diversity. 
Using custom annotation tooling we retopologise a head scan dataset such that eyelid creases are sharply defined and share a common edge-loop, with eyelid folds modeled explicitly. 
We retrain a parametric face model with this consistent data and demonstrate improved quality in 3D fitting, and accuracy in face shape reconstruction from images using multiple methods.

Our consistency definition assumes a single primary eyelid fold for all eyelids, though some individuals may have multiple folds or wrinkles on their eyelids. 
Our definition also assumes a common edge-loop for the epicanthal fold and eyelid fold, but these do not always connect in real eyelids. 
Future work involves broadening the consistency definition to accommodate representation of multiple eyelid folds and separate eyelid and epicanthal folds.

\appendix

\section{Diversity of Sampling}
\begin{figure}
    \centering
    \includegraphics[width=\linewidth]{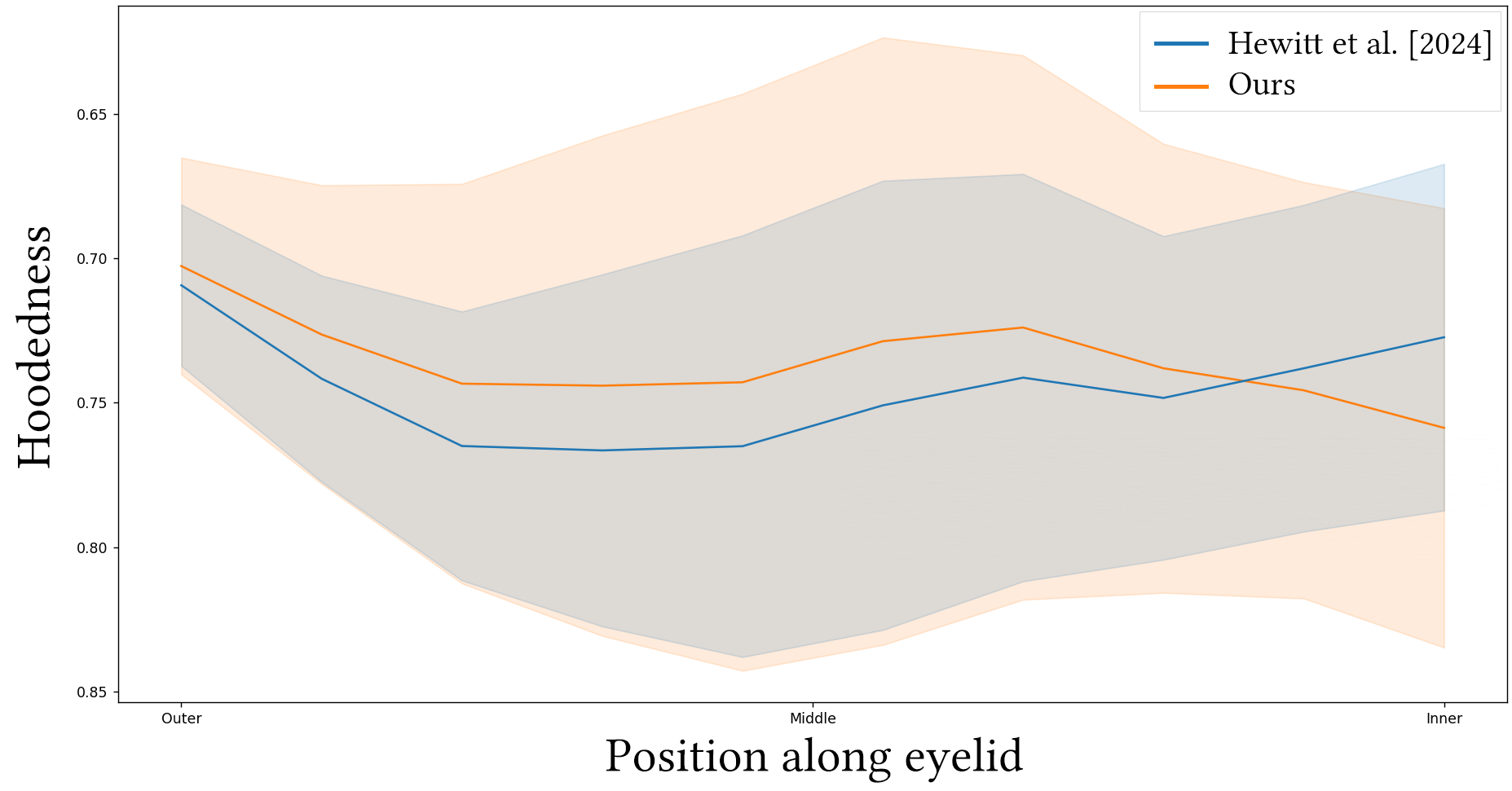}\\
    \vspace{-0.5em}
    \caption{Mean and standard deviation of eyelid profiles for 1000 random face samples drawn from a GMM fit to the training sets of \citet{hewitt2024look} (blue) and our updated model (orange). The orange areas at the top and bottom-right show our higher probability of sampling non-hooded eyelids and epicanthal folds.}
    \label{fig:eyelid-diversity}
\end{figure}

To compare diversity of the faces generated by each model, we measure the shape profiles of eyelids from 1000 GMM-sampled faces drawn from the models, and plot their mean and standard deviation in Figure~\ref{fig:eyelid-diversity}. 
The model trained on our data has significantly larger diversity than before, as the standard deviation is greater at all points on the eyelid fold. 
In particular, faces with non-hooded eyelids and epicanthal folds are more frequently sampled for our retrained model.

\section{Additional Results}

\begin{figure}
    \centering
    \footnotesize
    \begin{tabularx}{\linewidth}{YYY}
         Input Photo & \citet{hewitt2024look} & Ours
    \end{tabularx}
    \includegraphics[width=\linewidth]{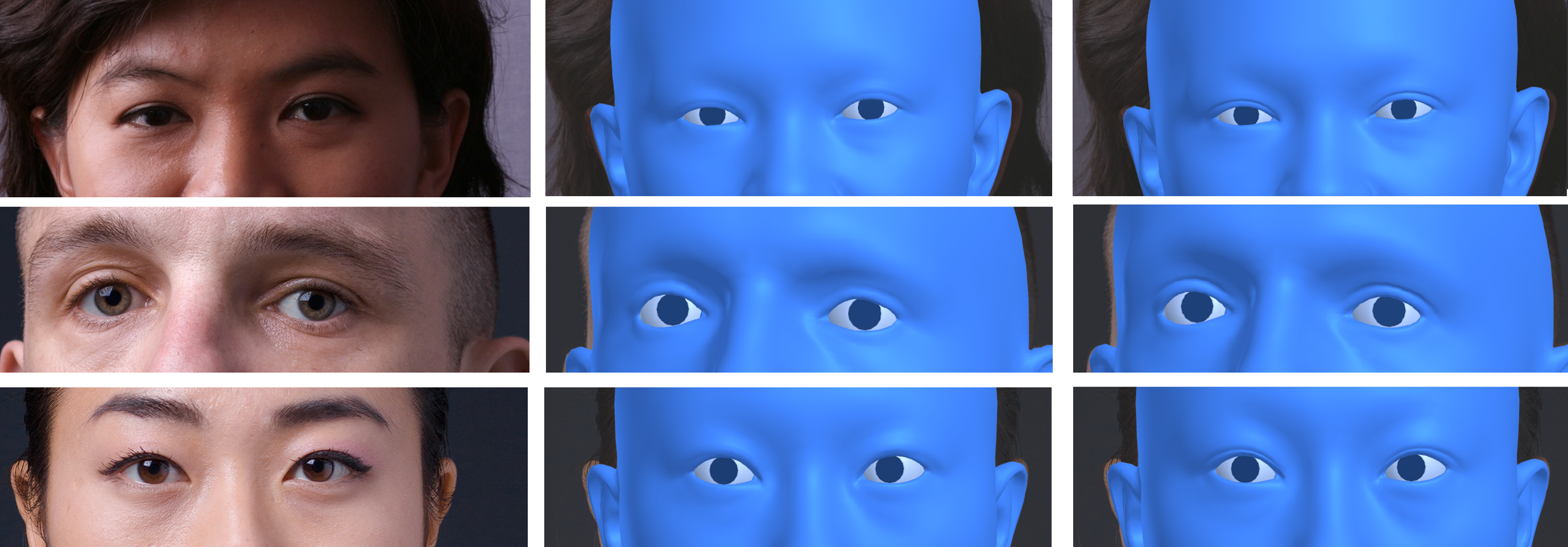}\\
    \vspace{-0.5em}
    \caption{Comparison of fits to 2D landmarks using the model of \citet{hewitt2024look} and our retrained model. \copylabel{Photos}{\citet{stirling}, Triplegangers}}
    \label{fig:registration-compare-more}
\end{figure}

\begin{figure}
    \centering
    \footnotesize
    \begin{tabularx}{\linewidth}{YYY}
         Input Photo & \citet{hewitt2024look} & Ours
    \end{tabularx}
    \includegraphics[width=\linewidth]{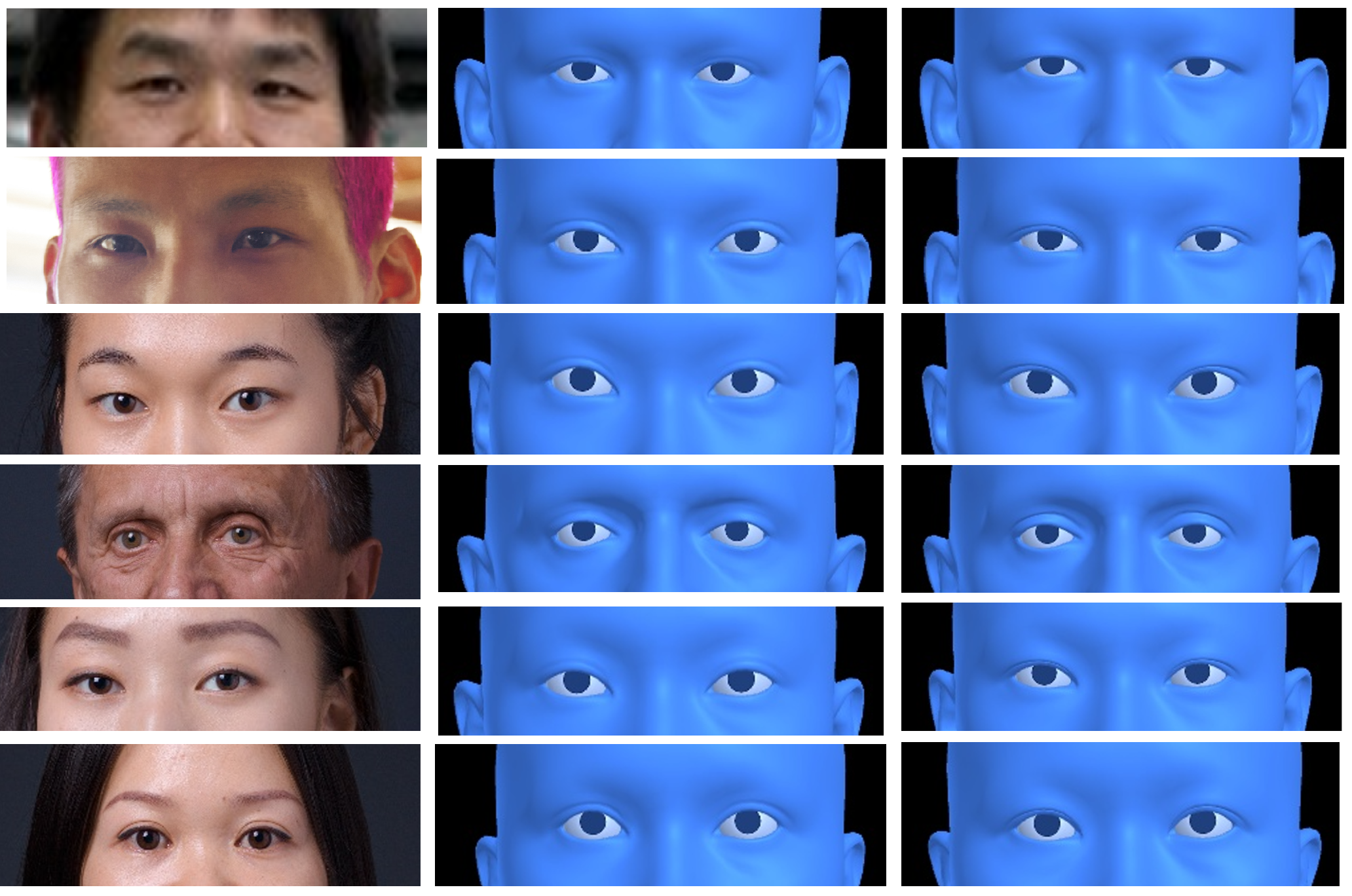}\\
    \vspace{-0.5em}
    \caption{Comparison of neural face shape prediction using the model of \citet{hewitt2024look} and our retrained model. \copylabel{Photos}{Microsoft, Triplegangers}}
    \label{fig:mifi-compare-more}
\end{figure}

Additional comparisons for landmark-based fitting results between our model and that of \citet{hewitt2024look} can be found in Figure~\ref{fig:registration-compare-more}, and for neural face shape prediction in Figure~\ref{fig:mifi-compare-more}.

\bibliographystyle{ACM-Reference-Format}
\bibliography{refs}

\end{document}